\documentclass[pra,aps,twocolumn,superscriptaddress]{revtex4}
\usepackage{hyperref}
\usepackage{bm}
\usepackage{epsfig}
\usepackage{amssymb}
\usepackage{graphicx}

\usepackage{amsmath}
\usepackage{epsfig}

\begin{document}
\title{Entanglement sharing with separable states}
\author{Ladislav Mi\v{s}ta Jr.}
\affiliation{Department of Optics, Palack\' y University, 17.
listopadu 12,  771~46 Olomouc, Czech Republic}

\date{\today}

\begin{abstract}
We propose an entanglement sharing protocol based on separable
states. Initially, two parties, Alice and Bob, share a two-mode
separable Gaussian state. Alice then splits her mode into two
separable modes and distributes them between two players. Bob is
separable from the players but he can create entanglement with
either of the players if the other player moves to his location
and collaborates with him. Any two parties are separable and the
creation of entanglement is thus mediated by transmission of a
mode which is separable from individual modes on Alice's and Bob's
side. For the state shared by the players and Bob one cannot
establish entanglement between any two modes even with the help of
operation on the third mode provided that Bob is restricted to
Gaussian measurements and the state thus carries a nontrivial
signature of bound entanglement. The present protocol also
demonstrates switching between different separability classes of
tripartite systems by coherent operations on its bipartite parts
and complements studies on protocols utilizing mixed partially
entangled multipartite states.

\end{abstract}

\maketitle
\section{Introduction}

Three correlated elementary quantum systems represent a basic
primitive which already may exhibit genuine multipartite
phenomena. The discovery that tripartite entanglement can provide
a stronger violation of local realism \cite
{Greenberger_89,Pan_00} than bipartite one has triggered a large
research activity with the aim to characterize it and find its
applications. Early studies of tripartite entanglement focused on
systems of three two-level particles (qubits) for which new forms
of multipartite bound entanglement \cite{Bennett_99a,Dur_00a} and
inequivalent entanglement classes \cite{Dur_00b} have been found.
Three qubits also proved to be a suitable platform for
construction of a classical analog of bound entanglement know as
bound information \cite{Acin_04}, which so far has not been found
in the bipartite scenario. In comparison with two qubits which can
only be separable or entangled three-qubit states can be divided
into five separability classes \cite{Dur_99} in dependence on
their separability properties with respect to different qubits.
Most of the applications of the three-qubit entanglement rely on
the utilization of pure states from the class of fully inseparable
states which are entangled with respect to all three qubits. They
involve various protocols for information splitting ranging from
secret sharing \cite{Hillery_99}, telecloning \cite{Murao_99} and
assisted teleportation \cite{Karlsson_98} as well as protocols for
construction of quantum gates \cite{Gottesman_99} or controlled
quantum cryptography \cite{Zukowski_98}.

Tripartite entanglement has also been investigated within the
framework of Gaussian states \cite{Braunstein_05} of
infinitely-dimensional quantum systems. A convenient prototype of
such a system is a system of three modes $A$, $A'$ and $B$ of an
electromagnetic field which are characterized by position $x_{j}$
and momentum $p_{j}$ quadrature operators satisfying the canonical
commutation rules $[x_j,p_k]=i\delta_{jk}$, $j,k=A,A',B$. Quantum
states of three modes can be represented in phase space by a
6-variate Wigner quasiprobability distribution \cite{Wigner_32}
and the set of Gaussian states comprises states with a Gaussian
Wigner function. A three-mode Gaussian state $\rho_{AA'B}$ is
therefore fully characterized by the vector of coherent
displacements $d=\langle\xi\rangle=\mbox{Tr}(\rho_{AA'B}\xi)$,
where we have introduced a column vector
$\xi=(x_A,p_A,x_{A'},p_{A'},x_B,p_B)^{T}$, and by a $6\times6$
covariance matrix (CM) with elements
$\gamma_{ij}=\langle\{\xi_i-d_i,\xi_j-d_j\}\rangle$,
$i,j=1,\ldots,6$, where $\{A,B\}=AB+BA$ is the anticommutator.

Following the classification of Ref.~\cite{Dur_99} we can divide
three-mode states into five separability classes
involving \cite{Giedke_01}:\\
1. {\it Fully inseparable states} which are entangled with respect
to all three bipartite splittings of modes $A,A'$ and $B$ into two
groups. That is states entangled across $A-(A'B)$, $A'-(AB)$ as
well as
$B-(AA')$ splitting.\\
2. {\it One-mode biseparable states} which are entangled with
respect to two bipartite splittings, but separable with respect to
the third one. Such a state exhibits entanglement across, e.g, $A-(A'B)$
and $A'-(AB)$ splitting but it is separable with respect to $B-(AA')$
splitting.\\
3. {\it Two-mode biseparable states} which are entangled across
one bipartite splitting, but separable with respect to the
remaining two splittings. The state is therefore entangled, e.g.,
across $A-(A'B)$ splitting but separable with respect to
$A'-(AB)$ and $B-(AA')$ splittings.\\
4. {\it Three-mode biseparable states} which are separable across
all three bipartite splittings but which cannot be
written as a convex mixture of triple product states.\\
5. {\it Fully separable states} which can be written as a convex
mixture of triple product states.

Like in the qubit case a genuine tripartite entanglement carried
by fully inseparable three-mode states is practically exclusively
used as a resource in quantum information protocols. It is due to
a relative ease of its preparation \cite{Loock_00,Aoki_03},
detection \cite{Loock_03} and a number of quantum protocols which
this type of entanglement offers
\cite{Loock_01,Yonezawa_04,Tyc_02}. On the other hand, the other
classes carry only partial or no entanglement, some exist just in
the mixed-state scenario (two-mode and three-mode biseparable
states \cite{Adesso_06}) and one might be then tempted to doubt
about their practical utility. However, the astonishing protocol
for entanglement distribution by a separable system
\cite{Cubitt_03} teaches us about the opposite. Originally
developed for qubits \cite{Cubitt_03}, later extended to Gaussian
states \cite{Mista_09}, and experimentally demonstrated in
\cite{Vollmer_13}, it shows that also other, even mixed and just
partially separable states may demonstrate new phenomena which are
not encountered in the context of fully inseparable states. It
demonstrates that two distant observers, Alice and Bob, can
entangle modes $A$ and $B$ held by them by sending a third
separable mode $A'$ between them. Initially, Alice holds modes $A$
and $A'$ whereas Bob holds mode $B$ of a suitable fully separable
Gaussian state. By a beam splitter on modes $A$ and $A'$ Alice
then transforms the state to the state entangled only across
$A-(A'B)$ splitting (two-mode biseparable state) and transmits the
separable mode $A'$ to Bob. He finally superimposes the received
mode $A'$ with his mode $B$ on another beam splitter and thus he
entangles modes $A$ and $B$ whereas mode $A'$ still remains
separable from the two-mode subsystem $(AB)$ (one-mode biseparable
state). The protocol thus shows that direct transmission of
entanglement is not necessary to entangle two separate parties but
only communication of a separable quantum system, local operations
and a priori shared suitable fully separable state of three
quantum systems suffice to accomplish the task.

In this paper we demonstrate the utility of mixed partially
separable states from other separability classes for
implementation of a certain entanglement sharing scheme.
Specifically, we show that there is a correlated separable
Gaussian state of two possibly distant modes $A$ and $B$ such that
if mode $A$ is split into two modes $A$ and $A'$ this creates
entanglement of each of the modes with the distant mode $B$ and
the other split mode taken together. More precisely, the splitting
entangles mode $A$ with the pair of modes $(A'B)$ and mode $A'$
with the pair of modes $(AB)$ at the same time. Moreover, any two
modes are separable in the prepared three-mode state. By
transmitting the mode $A'$ ($A$) which is {\it separable} from the
mode $A$ ($A'$) to the location of mode $B$ the entanglement can
be transformed by a simple beam splitter into entanglement of mode
$A$ ($A'$) and the distant mode $B$. The protocol thus shares
similarity with the quantum secret sharing protocol
\cite{Hillery_99,Cleve_99} for entanglement \cite{Choi_12} in
which a dealer splits one part of a bipartite entanglement among
several players in such a way that some collections of players can
recover the entanglement with the dealer whereas the other
collections cannot. If in the present case the modes $A$ and $B$
of the initial separable state are shared by two observers, called
Alice and Bob, then entanglement created by splitting on Alice's
side of mode $A$ into two modes $A$ and $A'$ can be turned into
the entanglement between Alice and Bob only if Bob has physically
at his disposal also either of the split modes $A$ or $A'$.
Remarkably, there is a certain interval of entanglement strengths
for which Bob can establish entanglement with Alice by the
coherent beam splitting operation on his mode and the received
mode but he cannot establish it by any Gaussian measurement on his
mode $B$ followed by displacement of modes $A$ and $A'$ which
challenges the question about the presence of bound entanglement
in the considered state. Like in the case of the entanglement
distribution by a separable ancilla the protocol works only with
mixed states and starts with a suitable fully separable three-mode
Gaussian state. In contrast with a two-mode biseparable state
which is created in the second step of the entanglement
distribution protocol a one-mode biseparable state is created in
the intermediate step of the present entanglement sharing
protocol. Likewise, a one-mode biseparable state appears at the
final step of the entanglement distribution protocol whereas the
entanglement sharing scheme is crowned by a fully inseparable
state.

The paper is structured as follows. In Sec~\ref{sec_1} we explain
the entanglement sharing protocol. In Sec.~\ref{sec_2} we show the
gap between unitary and measurement-based localizability of the
entanglement for the state from the intermediate step of the
protocol. Sec.~\ref{sec_3} contains discussion and conclusion.

\begin{figure}[tb]
\includegraphics[width=5.5cm,angle=90]{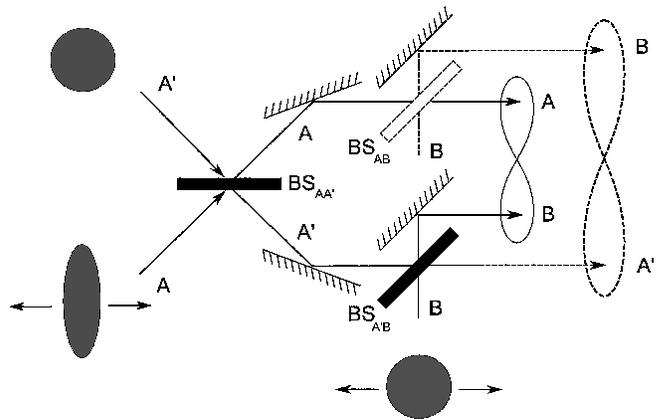}
\caption{Scheme of the entanglement sharing protocol. Mode $A$ in
a position squeezed vacuum state and a vacuum mode $B$ are
displaced as in Eq.~(\ref{displacements}). Mode $A$ is then split
on a balanced beam splitter $BS_{AA'}$ into two modes $A$ and $A'$
which creates a state with no two-mode entanglement which
possesses entanglement across $A-(A'B)$ and $A'-(AB)$ splittings
and which is separable across $B-(AA')$ splitting. If modes $A'$
and $B$ are superimposed on a balanced beam splitter $BS_{A'B}$
and the dashed beam splitter $BS_{AB}$ is absent, entanglement is
localized between modes $A$ and $B$ (solid lines with arrows). If
instead the beam splitter $BS_{A'B}$ is removed and modes $A$ and
$B$ are superimposed on a dashed balanced beam splitter $BS_{AB}$
entanglement is localized between modes $A'$ and $B$ (dashed lines
with arrows). In both cases the resulting state is entangled
across all three bipartite splittings and therefore the state is
genuinely tripartite entangled. See text for details.}\label{fig1}
\end{figure}

\section{Entanglement sharing protocol}\label{sec_1}

The scheme of the protocol is depicted in Fig.~\ref{fig1} The protocol
starts with a Gaussian state of two modes $A$ and $B$ shared by a
sender Alice and a receiver Bob which has the covariance matrix
(CM) of the form:
\begin{eqnarray}\label{gammaAB}
\gamma_{AB}=\left(\begin{array}{cccc}
1+e^{-2r}(e^{2\varepsilon}-1) & 0 & e^{-2r}-1 & 0 \\
0 & e^{2r} & 0 & 0\\
e^{-2r}-1 & 0 & 2-e^{-2r} & 0 \\
0 & 0 & 0 & 1 \\
\end{array}\right),
\end{eqnarray}
where $r\geq0$ is the squeezing parameter and $\varepsilon\geq0$
is a noise parameter which will be specified later. The right
upper $2\times2$ off-diagonal block is a diagonal matrix
$\mbox{diag}(e^{-2r}-1,0)$ with zero determinant and hence the
state is separable \cite{Simon_00}. The state can be easily
prepared by displacing position quadratures of the state in mode
$A$ with the CM
$\gamma_{A}=\mbox{diag}(e^{-2(r-\varepsilon)},e^{2r})$ and the
vacuum mode $B$ with CM $\gamma_{B}=\openone$ as
\begin{equation}\label{displacements}
x_{A}\rightarrow x_{A}+\bar{x},\quad x_{B}\rightarrow
x_{B}-\bar{x}.
\end{equation}
Here, $\bar{x}$ is the classical Gaussian distributed displacement
with variance satisfying $\langle\bar{x}^2\rangle=(1-e^{-2r})/2$
and $\openone$ is the $2\times2$ identity matrix.

Next, Alice splits her mode
$A$ into two modes $A$ and $A'$ by superimposing mode $A$ with
vacuum mode $A'$ on a balanced beam splitter. If we denote the
vacuum CM of mode $A'$ as $\gamma_{A'}=\openone$ and describe the
beam splitter by the orthogonal matrix
\begin{eqnarray}\label{Uij}
U_{ij}=\frac{1}{\sqrt{2}}\left(\begin{array}{cc}
\openone & \openone \\
\openone & -\openone\\
\end{array}\right),
\end{eqnarray}
where $i=A$, $j=A'$, we get the three-mode Gaussian state with the
following CM
\begin{eqnarray}\label{gammaAAprB}
\gamma_{AA'B}=\left(\begin{array}{ccc}
\alpha & \delta & \tau \\
\delta & \alpha & \tau\\
\tau & \tau & \beta\\
\end{array}\right),
\end{eqnarray}
with $\alpha,\beta,\tau$ and $\delta$ being diagonal matrices of
the form
$\alpha=\mbox{diag}(2+e^{-2r}(e^{2\varepsilon}-1),e^{2r}+1)/2$,
$\beta=\mbox{diag}(2-e^{-2r},1)$,
$\tau=\mbox{diag}(e^{-2r}-1,0)/\sqrt{2}$ and
$\delta=\mbox{diag}(e^{-2r}(e^{2\varepsilon}-1),e^{2r}-1)/2$.

The performance of the proposed protocol is enabled by the
remarkable separability properties of the state with CM
(\ref{gammaAAprB}). Let us investigate first the separability of
the state with respect to the splitting of modes $A,A'$ and $B$
into two groups ($1\times2$-mode separability). Beam splitting
transformation (\ref{Uij}) on mode $A$ and mode $A'$ cannot create
entanglement with mode $B$ and hence the state is separable across
the $B-(AA')$ splitting. The state is, however, entangled with
respect to the remaining $A-(A'B)$ and $A'-(AB)$ splittings as can
be easily proven using the positive partial transposition
criterion \cite{Peres_96,Horodecki_96} expressed in terms of
symplectic invariants \cite{Serafini_06}. Specifically,
separability of mode $X$ from a pair of modes $(YZ)$ in a generic
Gaussian state of three modes $X,Y$ and $Z$ with the CM $\gamma$
can be determined from the symplectic invariants of the matrix
$\gamma^{(T_x)}\equiv\Lambda_{x}\gamma\Lambda_{x}^{T}$ associated
with the partial transpose of the state with respect to mode $X$.
Here,
$\Lambda_{x}\equiv\sigma_{z}^{(X)}\oplus\openone^{(Y)}\oplus\openone^{(Z)}$,
where $\sigma_{z}=\mbox{diag}(1,-1)$ is the Pauli diagonal matrix.
The matrix $\gamma^{(T_x)}$ possesses three symplectic invariants
denoted as $I_1,I_2$ and $I_3=\mbox{det}(\gamma)$ which coincide
with the coefficients of the characteristic polynomial of the
matrix $\Omega\gamma^{(T_{x})}$, i.e.
\begin{equation}\label{polynomial}
\mbox{det}(\Omega\gamma^{(T_{x})}-q\openone)=q^{6}+I_1q^{4}+I_2q^{2}+I_3,
\end{equation}
where
\begin{eqnarray}\label{Omega}
\Omega=\bigoplus_{i=1}^{3}\left(\begin{array}{cc}
0 & 1 \\
-1 & 0\\
\end{array}\right).
\end{eqnarray}
According to the criterion \cite{Serafini_06} mode $X$ is
entangled with the two-mode subsystem $(YZ)$ if
\begin{equation}\label{Sigmadef}
\Sigma_{x}=I_3-I_2+I_1-1<0.
\end{equation}
In the case of the CM (\ref{gammaAAprB}) one gets explicitly after some algebra that
\begin{equation}\label{Sigma}
\Sigma_{A}=8e^{\varepsilon-r}\sinh(\varepsilon-r)\sinh^2(r),
\end{equation}
which implies that if $r>\varepsilon$ it holds that $\Sigma_{A}<0$
and there is entanglement across $A-(A'B)$ splitting. Owing to the
symmetry of the state under the exchange of modes $A$ and $A'$
(bisymmetric state \cite{Serafini_05}) it follows immediately that
$\Sigma_{A}=\Sigma_{A'}$ and the state is therefore also entangled
across $A'-(AB)$ splitting. Thus, the studied state is separable
across one bipartite splitting and therefore it belongs to the
class of one-mode biseparable Gaussian states.

Let us focus now on the separability of the two-mode reductions
($1\times1$-mode separability) of the state with the CM
(\ref{gammaAAprB}). Due to the separability of the $B-(AA')$
splitting mode $B$ is inevitably separable both from the mode $A$
as well as from the mode $A'$. The reduced state of modes $A$ and
$A'$ was created by mixing of the state with CM $\gamma_{A}=
\mbox{diag}[1+e^{-2r}(e^{2\varepsilon}-1),e^{2r}]$ with the vacuum
state on a beam splitter (\ref{Uij}). Both the eigenvalues of the
CM $\gamma_{A}$ are lower bounded by the unity and hence the
corresponding normally ordered CM
$\gamma_{A}^{(\mathcal{N})}\equiv\gamma_{A}-\openone$ is positive
semidefinite. The normally ordered characteristic function of the
state then possesses a Fourier transform which is not more
singular than a Dirac delta function and the state with the CM
$\gamma_{A}$ is thus classical. As mixing of such a state with a
vacuum state on a beam splitter cannot create entanglement
\cite{Kim_02} mode $A$ is therefore separable from mode $A'$ in
the state with CM (\ref{gammaAAprB}). In summary, for the state
with the CM (\ref{gammaAAprB}) any two modes are separable.

Note, that the aforementioned separability properties of the state
with CM (\ref{gammaAAprB}) can exist only in a mixed-state
scenario. Indeed, for a pure state separability of mode $B$ from
modes $(AA')$ implies the state to be a product state across the
$B-(AA')$ splitting. Likewise, the separability of the mode $A$
from the mode $A'$ implies that the reduced state of the modes $A$
and $A'$ is also a product state. Consequently, a pure three-mode
state where the mode $B$ is separable from modes $(AA')$ and mode
$A$ is at the same time separable from mode $A'$ is therefore a
triple product state which is fully separable. Such a state,
however, cannot possess entanglement across, e.g., $A-(A'B)$
splitting as is the case of our state.

At the final stage of the protocol Alice keeps mode $A$ and sends
the separable mode $A'$ to Bob. He superimposes the mode with his
mode $B$ on a balanced beam splitter $U_{BA'}$ given in
Eq.~(\ref{Uij}) where $i=B$ and $j=A'$ which creates a state with CM
\begin{eqnarray}\label{tildegammaAAprB}
\tilde{\gamma}_{AA'B}=\left(\begin{array}{ccc}
\alpha & \frac{\tau-\delta}{\sqrt{2}} & \frac{\tau+\delta}{\sqrt{2}} \\
\frac{\tau-\delta}{\sqrt{2}} & \frac{\alpha+\beta-2\tau}{2} & \frac{\beta-\alpha}{2}\\
\frac{\tau+\delta}{\sqrt{2}} & \frac{\beta-\alpha}{2} & \frac{\alpha+\beta+2\tau}{2}\\
\end{array}\right),
\end{eqnarray}
where the $2\times 2$ submatrices $\alpha,\beta,\tau$ and $\delta$
are given below Eq.~(\ref{gammaAAprB}). If, on the other hand, Alice keeps mode $A'$ and sends the separable mode $A$ to Bob who
mixes it with his mode on the beam splitter
\begin{eqnarray}\label{UAB}
U_{AB}=\frac{1}{\sqrt{2}}\left(\begin{array}{cc}
\openone & -\openone \\
\openone & \openone\\
\end{array}\right),
\end{eqnarray}
the CM of the resulting state reads
\begin{eqnarray}\label{tildetildegammaAAprB}
\tilde{\tilde{\gamma}}_{AA'B}=\left(\begin{array}{ccc}
\frac{\alpha+\beta-2\tau}{2} & \frac{\delta-\tau}{\sqrt{2}} & \frac{\alpha-\beta}{2}\\
\frac{\delta-\tau}{\sqrt{2}} & \alpha & \frac{\delta+\tau}{\sqrt{2}} \\
\frac{\alpha-\beta}{2} & \frac{\delta+\tau}{\sqrt{2}} &  \frac{\alpha+\beta+2\tau}{2}\\
\end{array}\right).
\end{eqnarray}

By retaining only the modes $A$ ($A'$) and $B$ it then follows that Alice and Bob are left with a reduced
two-mode state with the CM
\begin{eqnarray}\label{tildegammaAB}
\tilde{\gamma}_{AB}=\tilde{\tilde{\gamma}}_{A'B}=\left(\begin{array}{cc}
\alpha & \frac{\delta+\tau}{\sqrt{2}} \\
\frac{\delta+\tau}{\sqrt{2}} & \frac{\alpha+\beta+2\tau}{2}\\
\end{array}\right),
\end{eqnarray}
which describes an entangled state provided that the squeezing parameter $r$ is large enough.
The threshold squeezing $r_{\rm e}$ above which the entanglement appears can be derived from
the two-mode version of the sufficient condition for entanglement \cite{Serafini_06}
\begin{equation}\label{Sigma2}
\mbox{det}\tilde{\gamma}_{AB}-\Delta+1<0,
\end{equation}
where $\Delta=\mbox{det}\alpha+\frac{1}{4}\mbox{det}\left(\alpha+\beta+2\tau\right)-\mbox{det}\left(\delta+\tau\right)$.
Substituting here for the matrices $\alpha,\beta,\tau$ and $\delta$ defined below
Eq.~(\ref{gammaAAprB}) one finds, that mode $A$ ($A'$) is entangled with mode $B$ if the
squeezing $r$ satisfies $r>r_{\rm e}$, where
\begin{widetext}
\begin{eqnarray}\label{re}
r_{\rm e}=\frac{1}{2}\ln\left[\frac{11e^{2\varepsilon}+8\sqrt{2}-13+
\sqrt{(11e^{2\varepsilon}+8\sqrt{2}-13)^2+4e^{2\varepsilon}(8\sqrt{2}-1)}}{2(8\sqrt{2}-1)}\right].
\end{eqnarray}
\end{widetext}

It is further of interest to look at the $1\times 2$-mode
separability of the states with CMs (\ref{tildegammaAAprB}) and
(\ref{tildetildegammaAAprB}) from the last step of the protocol.
We have already seen that for the state with the CM
(\ref{tildegammaAAprB}) [(\ref{tildetildegammaAAprB})] mode $A$
($A'$) is entangled with mode $B$. This implies, that there is
entanglement across $A-(A'B)$ $[A'-(AB)]$ as well as $B-(AA')$
$[B-(AA')]$ splitting. But what about entanglement across the
remaining $A'-(AB)$ $[A-(A'B)]$ splitting? Analyzing the
entanglement using again the criterion (\ref{Sigmadef}) one gets
for the CMs (\ref{tildegammaAAprB}) and
(\ref{tildetildegammaAAprB}) the following expressions:
\begin{equation}\label{tildeSigma}
\tilde{\Sigma}_{A'}=\tilde{\tilde{\Sigma}}_{A}=\frac{\Sigma_{A}}{4},
\end{equation}
where the quantity $\tilde{\Sigma}_{A'}$
($\tilde{\tilde{\Sigma}}_{A}$) characterizes separability of the
mode $A'$ ($A$) in the state with the CM (\ref{tildegammaAAprB})
[(\ref{tildetildegammaAAprB})] and $\Sigma_{A}$ is given in
Eq.~(\ref{Sigma}). Hence, for the considered squeezing
$r>\varepsilon$ the state in the last step of the present protocol
is in both cases entangled across all three bipartite splittings
and it therefore carries a genuine three-mode entanglement. The
proposed sharing scheme thus also illustrates remarkable
transformation properties of the three-mode fully separable state
given by a product of a separable state of modes $A$ and $B$ with
CM (\ref{gammaAB}) and a vacuum state of mode $A'$. Namely, the
beam splitting transformation (\ref{Uij}) on a two-mode subsystem
formed by modes $A$ and $A'$ transforms the state into the
one-mode biseparable state which is separable with respect to
$B-(AA')$ splitting and entangled across $A-(A'B)$ and $A'-(AB)$
splittings. Moreover, the second beam splitter on modes $A'$ ($A$)
and $B$ preserves the latter entanglement and further creates
entanglement also across $B-(AA')$ splitting, i.e., creates a
genuine three-mode entanglement.

\section{Unitary versus measurement-based localizability of the
intermediate entanglement}\label{sec_2}

It can seem for the first sight that Bob may not need coherent
beam splitting operation on his mode $B$ and the received mode
$A'$ ($A$) to establish entanglement with Alice's mode $A$ ($A'$).
One can argue that the participants could first establish
entanglement between the transmitted mode $A'$ ($A$) (held by Bob)
and Alice's mode $A$ ($A'$) simply by optimally measuring mode $B$
(which is separable from the pair of modes $(AA')$) followed by an
optimal displacement of modes $A$ and $A'$. The entanglement thus
obtained could be subsequently transformed into the entanglement
of mode $B$ with Alice's mode $A$ ($A'$) just by swapping the
transmitted mode $A'$ ($A$) with the mode $B$. Now we show, that
if Bob is restricted to Gaussian measurements there is a region of
squeezing parameters $r$ for which he is unable to establish any
entanglement with Alice by this measure-and-displace strategy. The
task to be solved can be formulated as a task for finding maximum
entanglement which can be localized between modes $A$ and $A'$ of
the state with CM (\ref{gammaAAprB}) by optimal Gaussian
measurement on mode $B$ of the state. This problem, known as
Gaussian localizable entanglement, has already been solved in the
literature \cite{Fiurasek_07}. It can be conveniently analyzed
using the concept of lower symplectic eigenvalue of the partially
transposed state, which is for a generic CM $\sigma_{AB}$ of two
modes $A$ and $B$ written in the $2\times2$-block form with
respect to $A-B$ splitting
\begin{eqnarray}\label{sigmaAB}
\sigma_{AB}=\left(\begin{array}{cc}
A & C \\
C^{T} & B\\
\end{array}\right),
\end{eqnarray}
given by \cite{Vidal_02}
\begin{eqnarray}\label{mu}
\mu=\sqrt{\frac{\Delta-\sqrt{\Delta^2-4\mbox{det}\sigma_{AB}}}{2}},
\end{eqnarray}
where $\Delta=\mbox{det}A+\mbox{det}B-2\mbox{det}C$. The state
with CM $\sigma_{AB}$ contains entanglement if and only if
$\mu<1$. The symplectic eigenvalue $(\ref{mu})$ also characterizes
the amount of entanglement in the state which can be quantified by
the logarithmic negativity
$E_{\mathcal{N}}(\sigma_{AB})=\mbox{max}(0,-\log_{2}\mu)
$\cite{Vidal_02}. As $E_{\mathcal{N}}$ is a monotonically
decreasing function of the symplectic eigenvalue it follows that
the smaller the value of $\mu$ the larger the entanglement. For
certain families of three-mode Gaussian states one can find even
analytically the symplectic eigenvalue (\ref{mu}) for a two-mode
state obtained by Gaussian measurement on the third mode minimized
over all Gaussian measurements on the mode \cite{Fiurasek_07}. The
present state with CM (\ref{gammaAAprB}) belongs to the class of
the bisymmetric states for which the problem of the Gaussian
localizable entanglement can also be fully resolved using the
analytical tools \cite{Mista_08}. It turns, that the localizable
entanglement between modes $A$ and $A'$ is always achieved by
homodyne detection of position quadrature $x_{B}$ on mode $B$,
i.e., by projection of the mode onto an infinitely squeezed
position eigenstate. This gives the minimal lower symplectic
eigenvalue (\ref{mu}) of the partial transpose of the conditional
state of modes $A$ and $A'$ in the form:
\begin{widetext}
\begin{eqnarray}\label{mum}
\mu_{\rm m}= \left\{
    \begin{array}{ll} e^{r}  & \mbox{if   } r<r_{\rm l}\equiv\frac{1}{2}\ln\left\{\frac{1}{3}+2\sqrt{-\frac{p}{3}}\cos\left[\frac{1}{3}\arccos\left(-\frac{q}{2}\sqrt{-\frac{27}{p^3}}\right)\right]\right\};\\
        \sqrt{1+e^{-2r}(e^{2\varepsilon}-1)-\frac{(e^{-2r}-1)^2}{2-e^{-2r}}} & \mbox{otherwise},
    \end{array}
\right.\nonumber\\
\end{eqnarray}
\end{widetext}
where
\begin{equation}\label{pq}
p=\frac{1}{6}-e^{2\varepsilon},\quad q=\frac{5}{54}+\frac{e^{2\varepsilon}}{6}.
\end{equation}
Clearly, the condition on the squeezing $r$ for which Bob can
localize entanglement between modes $A$ and $A'$ can be
derived by solving the inequality $\mu_{\rm m}<1$ with respect
to $r$. This gives explicitly that the localizable entanglement is
nonzero if $r>r_{\rm m}$, where
\begin{eqnarray}\label{rm}
r_{\rm m}=\frac{1}{2}\ln\left[e^{2\varepsilon}+\sqrt{e^{2\varepsilon}\left(e^{2\varepsilon}-1\right)}\right].
\end{eqnarray}

\begin{figure}[tb]
\includegraphics[width=8.5cm]{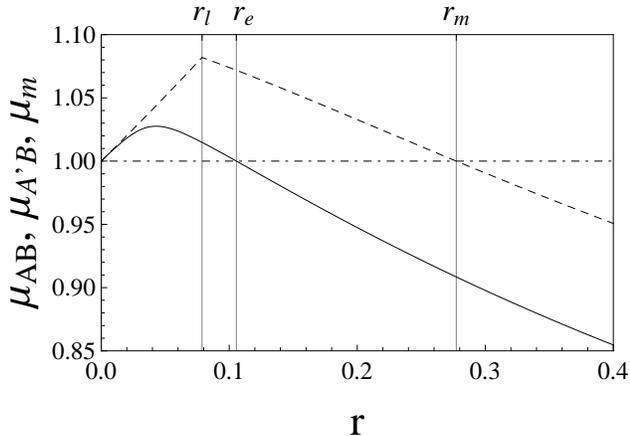}
\caption{Lower symplectic eigenvalue $\mu_{AB}=\mu_{A'B}$ of the
partial transpose of the states with CMs (\ref{tildegammaAB}) (solid
curve) and the symplectic eigenvalue $\mu_{\rm m}$,
Eq.~(\ref{mum}), (dashed curve) versus the squeezing parameter $r$
for $\varepsilon=0.1$. The solid vertical lines correspond to the
threshold squeezings $r_{\rm l}=0.079$, Eq.~(\ref{mum}),
$r_{\rm e}$, Eq.~(\ref{re}), and $r_{\rm m}$, Eq.~(\ref{rm}).
All the quantities plotted are dimensionless.}\label{fig2}
\end{figure}

\begin{figure}[tb]
\includegraphics[width=8.5cm]{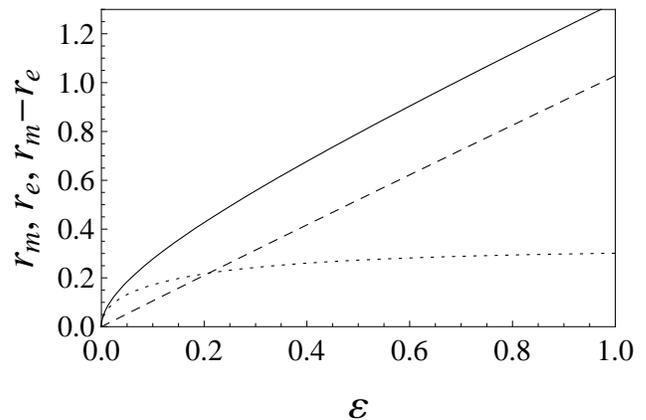}
\caption{Squeezing thresholds $r_{\rm m}$, Eq.~(\ref{rm}), (solid
curve), $r_{\rm e}$, Eq.~(\ref{re}), (dashed curve), and their
difference $r_{\rm m}-r_{\rm e}$ (dotted curve). All the
quantities plotted are dimensionless.}\label{fig3}
\end{figure}

The performance of our protocol can be illustrated on a particular
example when we set $\varepsilon=0.1$, which is depicted in
Fig.~\ref{fig2}. It follows from Eq.~(\ref{Sigma}) that the state
with the CM (\ref{gammaAAprB}) contains entanglement across $A-(A'B)$
as well as $A'-(AB)$ splittings if $r>\varepsilon=0.1$. The entanglement
can be transformed into two-mode entanglement of modes  $A$ ($A'$) and $B$
by a balanced beam splitter on modes $A'$ ($A$) and $B$ if $r>r_{\rm
e}\doteq0.106$. In contrast, Bob can localize some entanglement
between modes $A$ and $A'$ only if $r>r_m\doteq0.277$ according to
Eq.~(\ref{rm}). Thus, in the interval $r_m\geq r>r_{\rm
e}$ Bob can get entanglement by a coherent operation on his mode and the
received mode but cannot create entanglement between Alice's mode
and the received mode by any Gaussian measurement on his mode and
optimal displacement of Alice's and the received mode.
Further analysis shows that the gap between the threshold
squeezings (\ref{re}) and (\ref{rm}) exists also for the other
values of the parameter $\varepsilon$. In Fig.~\ref{fig3} we plot
the threshold squeezings as well as their difference $r_{\rm
m}-r_{\rm e}$ as a function of $\varepsilon$. The figure and
further calculations reveal that the difference exists for any
$\varepsilon>0$ and it is a monotonically increasing function of
$\varepsilon$ which asymptotically approaches the limit value
$(1/2)\ln[2(8\sqrt{2}-1)/11]\doteq0.314$. Thus, for the state with
the CM (\ref{gammaAAprB}) for any $\varepsilon>0$ there exists a
region of squeezings $r_{\rm m}\geq r>r_{\rm e}$ for which
entanglement between Alice and Bob cannot be established by
performing any Gaussian measurement on mode $B$ and optimally
displacing modes $A$ and $A'$ (where one of the modes is held by
Bob), but it can be created by superimposing Bob's mode $B$ and
the received mode (either $A'$ or $A$) on a balanced beam
splitter.

Note, that the gap exists also in the protocol for the Gaussian
entanglement distribution by a separable ancilla \cite{Mista_09}.
Here, the possibility to localize entanglement between mode $A$
and the transmitted mode $A'$ by measurement on Bob's mode $B$ is
prevented by separability of mode $A'$ from the pair of modes
$(AB)$. Nevertheless, a balanced beam splitter on modes $A'$ and
$B$ creates entanglement between modes $A$ and $B$. Passive
coherent unitary operations exhibit superiority over the method
based on the measurement and feed-forward also in the two-mode
scenario \cite{Filip_10}. In this case, the coherent operations
allow to extract squeezing from a squeezed signal classically
correlated to a probe in cases when the measurement on the probe
followed by a feed-forward correction on the signal fails.

\section{Discussion and conclusion}\label{sec_3}

The key state of our sharing protocol with the CM
(\ref{gammaAAprB}) is interesting from the point of view of the
nondistillable (bound) entanglement \cite{MHorodecki_98}. A
bipartite quantum state is nondistillable if it is impossible to
transform many copies of the state by local operations and
classical communication (LOCC) into fewer copies of nearly
maximally entangled singlet state. The concept of bound
entanglement can be easily generalized to three parties
\cite{Dur_00a} in analogy with the concept of the classical
multipartite bound information \cite{Acin_04}. We say that a
tripartite quantum state is bound entangled if i) any two parties
cannot distill singlet states by LOCC even with the help of the
third party and ii) the state cannot be created by LOCC. Examples
of the tripartite bound entanglement can be found both for
two-level systems (qubits) \cite{Bennett_99a,Dur_99} and Gaussian
states \cite{Giedke_01}. They are given by the two-mode
biseparable states which are separable across two bipartite
splittings and entangled across the third one as well as
three-mode biseparable states which are separable across all three
bipartite splittings but they are not fully separable. This is
because separability across at least two splittings immediately
guarantees satisfaction of the condition i) whereas the presence
of entanglement causes that also condition ii) holds. However,
inseparability with respect to at least two bipartite splittings
is currently known to be only a necessary condition for
distillability but it is not known whether it is also sufficient.
Thus although there are, for instance, distillable one-mode
biseparable states both in the qubit \cite{Dur_99,Cubitt_03} as
well as Gaussian case \cite{Mista_08} one cannot rule out the
possibility that there also exist bound entangled states belonging
to this class \cite{Dur_99,Giedke_01}. Our state with the CM
(\ref{gammaAAprB}) is entangled and hence fulfils the condition
ii) but moreover it also possesses nontrivial properties which are
necessary to satisfy the condition i). First, for any Gaussian
state satisfying i) any two parties have to be separable. Namely,
if there were entanglement between some pair of parties, then it
would be distillable \cite{Giedke_01b}. For our state any two
modes are separable and thus no entanglement can be distilled
between them if the third party is totally ignored. What is more,
not only entanglement cannot be distilled between $A$ and $B$ with
the help of $A'$ as well as between $A'$ and $B$ with the help of
$A$ because $B$ is separable from $(AA')$ but for certain
squeezings entanglement also cannot be distilled between $A$ and
$A'$ with the help of Bob provided that he is restricted to the
Gaussian measurements on his mode, which is a nontrivial necessary
condition for fulfillment of the requirement i).

In conclusion, we have constructed a two-mode separable Gaussian
state which can be transformed by splitting of one of its modes on
a beam splitter to a state without any two-mode entanglement but
with entanglement across two bipartite splittings. If the initial
two-mode state is shared by distant Alice and Bob, the beam
splitter on Alice's side creates entanglement between one of its
outputs and a non-local composite system composed of the other
output of the beam splitter and the distant Bob's mode. Although
the entanglement is between Alice's mode and the system involving
distant Bob's mode it is not entanglement between Alice and Bob
because Bob's mode is separable at the same time. The entanglement
can nevertheless be turned into entanglement between Alice and Bob
by sending one output mode of Alice's beam splitter (which is
separable from the other output mode) to Bob and mixing it with
Bob's mode on another beam splitter. The protocol just described
thus can be interpreted as an entanglement sharing scheme in which
entanglement created by Alice's beam splitter can be transformed
into entanglement with Bob only if Bob has at his disposal
physically also one output mode of the beam splitter. Besides, our
analysis shows that for one copy of the key one-mode biseparable
state of our protocol entanglement cannot be distilled between any
two modes with the help of the third party if Bob is restricted to
Gaussian measurements which is a nontrivial property necessary for
the presence of the bound entanglement in the state. The question
of whether the same holds true also for more generic even
non-Gaussian operations on Bob's mode and multiple copies is left
for further research. We believe that our results contribute to
the better understanding and utilization of entanglement as well
as separable correlations in multipartite mixed quantum states.

\section{Acknowledgment}

I would like to thank V. Chille, Ch. Peuntinger, Ch. Marquardt and
N. Korolkova for fruitful discussions. The research has been supported
by the GACR Project No. P205/12/0694.

\end{document}